\pdfoutput=1
\documentclass[aps,twocolumn,pra]{revtex4-1}
\usepackage{newlfont}
\usepackage{amssymb}
\usepackage{amsfonts}
\usepackage{amsmath}
\usepackage{wasysym}
\usepackage{graphicx}
\usepackage{bm}

\usepackage{epsfig}

\usepackage{amsthm}

\usepackage{times}

\begin{document}

\title{Violation of Invariance of Entanglement Under Local ${\cal {PT}}$ Symmetric Unitary}
%\title{Conditions for Monogamy of Quantum Correlations:\\ Monogamous Greenberger-Horne-Zeilinger versus Polygamous W states}
%\title{Conditions for Monogamy of Quantum Correlations:\\ Polygamous W vs. Monogamous GHZ}
%\title{Conditions of Monogamy of Quantum Correlations: Polygamous W vs. Monogamous GHZ states}
%\title{Monogamy of Quantum Correlations as Detector of Genuine Tripartite Entangled States}

\author{Arun Kumar Pati}

\affiliation{Quantum Information and Computation Group, \\
Harish-Chandra Research Institute, Chhatnag Road, Jhunsi, Allahabad 211 019, India}

\begin{abstract}

Entanglement is one of the key feature of quantum world and 
any entanglement measure must satisfy some basic laws. Most important of them is the
%non-increase of entanglement under local operations, and in particular, it is invariant 
invariance of entanglement under local unitary operations. We show that this is no longer true with local 
${\cal {PT}}$ symmetric unitary operations. If two parties share a maximally entangled state, 
then under local ${\cal {PT}}$ symmetric unitary evolution the entropy
of entanglement for pure bipartite states does not remain invariant.
%This proves our earlier conjecture that even though gloabl ${\cal {PT}}$ symmetry is equivalent to
%conventional quantum theory, local ${\cal {PT}}$ symmetry can have inequivalent predictions.  
Furthermore, we show that if one of the party has access to ${\cal {PT}}$-symmetric quantum world, then
a maximally entangled state in usual quantum theory appears as a non-maximally entangled states for the other party. 
This we call as the ``entanglement mismatch'' effect which can lead to the violation of the no-signaling condition.
\end{abstract}
%\date{\today}

\maketitle

\def\ra{\rangle}
\def\la{\langle}
\def\ver{\arrowvert}

%\pacs{03.67.-a, 03.65.Bz, 03.65.Ta, 03.67.Lx}

%]

%{\em Introduction.--}
Quantum theory is arguably the most fundamental theory of Nature which has been tested 
over more than hundred years. Though, there have been several attempts to extend quantum 
theory, the basic tenets of the theory remained untouched so far. There have been non-linear generalizations 
of quantum theory \cite{sw,sw1,gisin}, non-unitary modifications to the Schr{\"o}dinger equations \cite{ghi,diosi,gisin1,pen}, 
complex extension of
quantum theory \cite{bend1,bend2}, and many more, to name a few. However, these 
generalizations are not free from importunate issues.

In conventional quantum theory the 
observables are represented by Hermitian operators and the evolution of 
a closed system is governed by unitary evolution. However, 
in recent years there have been considerable interests in quantum systems governed by
non-Hermitian Hamiltonians \cite{bend1,bend2,bend3,most,most1,most2,most3}.
It was discovered that there are class of non-Hermitian Hamiltonians which
posses real eigenvalues provided they respect 
${\cal {PT}}$ symmetry and the symmetry is unbroken. 
In ${\cal {PT}}$-symmetric quantum mechanics the usual condition 
of Hermiticity of
operators is replaced by the condition of ${\cal {CPT}}$ invariance, 
where ${\cal C}$ stands 
for conjugation, ${\cal P}$ for parity and ${\cal T}$ for time reversal \cite{bend1}. 
In standard quantum theory ${\cal {CPT}}$ symmetry and Hermiticity 
conditions are the same. The ${\cal {CPT}}$ invariance condition is a natural 
extension of 
Hermiticity condition that allows reality of observables and unitary dynamics.
Using the operator ${\cal C}$, Bender et al \cite{bend2} have introduced an inner 
product structure 
associated with ${\cal {CPT}}$ which can have positive definite norms for quantum states.
%It may be mentioned that the ${\cal {CPT}}$ inner product was also discovered in 
%Ref. \cite{nev}.

Entanglement is one of the weirdest feature of quantum mechanics. 
In the emerging field of quantum information theory
entanglement plays a major role \cite{horo}. This is also a very useful resource in the 
sense that using entanglement one can do many things in the quantum world
which are usually impossible in ordinary classical world. Some of these tasks 
include, but not limited to, quantum computing \cite{qc}, quantum teleportation \cite{qt}, 
quantum cryptography \cite{qcr}, remote state preparation \cite{akp},
and quantum communication \cite{quc}. Usual discussions about quantum entanglement pertain to the realm of
Hermitian quantum theory. However, recently the notion of entanglement for quantum systems 
described by ${\cal {PT}}$ symmetric Hamiltonians was 
introduced by the present author \cite{pati} and independently in Ref. \cite{sola}.

Early formulation 
of ${\cal {PT}}$-symmetric quantum theory aimed to offer a genuine extension of 
usual quantum theory. Later, mathematical unitary equivalence has been 
shown between ${\cal {PT}}$ symmetric quantum theory and the usual quantum theory for single quantum
systems \cite{most}. However, entangled quantum systems may offer new 
insights into 
the nature of this (in)equivalence. Since the equivalence properties of 
entangled states are 
different under joint unitary and under local unitary transformations, it was 
conjectured by the present author that under local unitary transformations 
(or more generally under LOCC paradigm) equivalence between ${\cal {PT}}$ symmetric 
quantum theory and 
the usual quantum theory may not exists \cite{pati}. 

In addition, ${\cal {PT}}$ symmetric quantum theory offers new possibilities.
It has been shown that faster time evolution \cite{bend4} and state discrimination is possible with 
${\cal {PT}}$ symmetric Hamiltonian \cite{bend5}. Though, the faster time evolution with non-Hermitian Hamiltonian 
has been questioned in Ref. \cite{most4}.
One assumption that is usually made 
is that one can describe a local subsystem by a ${\cal {PT}}$ symmetric Hamiltonian and it 
is possible to switch between ${\cal {PT}}$ symmetric world and the conventional quantum world.
%This assumption (which is stated as two separate assumptions) has been recently scrutinized, 
Similar assumptions \cite{sam,sams} have been recently scrutinized, 
and it has been shown that local ${\cal {PT}}$ symmetry acting on a composite system can lead to signaling \cite{lee}.

%We have introduced the notion of ${\cal {PT}}$-symmetric quantum bit (PTQubit). 

Any entanglement measure must satisfy some basic laws. Most important of them is the
non-increase of entanglement under local operations.  Moreover, a stringent requirement is that it must be invariant 
under local unitary operations. In this paper, we show that this does not hold for local 
${\cal {PT}}$ symmetric unitary operations. If Alice and Bob share a maximally entangled state 
then under local ${\cal {PT}}$ symmetric unitary evolution the entropy 
 of entanglement for a pure bipartite states does not remain invariant.
%Furthermore, we will show that maximally entangled states in usual quantum 
%theory behave like non-maximally entangled states in ${\cal {PT}}$-symmetric 
%quantum theory. 
This proves our earlier conjecture that even though global ${\cal {PT}}$ symmetry is equivalent to
conventional quantum theory, local ${\cal {PT}}$ symmetry can have inequivalent predictions. 
Because of the ${\cal {CPT}}$ inner product, orthogonal 
quantum states in ordinary quantum theory become non-orthogonal quantum 
states in non-Hermitian quantum theory. 
As a consequence, we will show that a maximally entangled state in ordinary 
theory can appear as a non-maximally entangled state to one observer if another observer has access to 
${\cal {PT}}$ symmetric quantum world. This is a precursor for the violation of the no-signaling condition.

${\cal {PT}}$ {\em Symmetric Quantum Theory.--}
We will give the basic formalism that is necessary to develop
the notion of entanglement in non-Hermitian quantum theory.
In earlier formulation of 
${\cal {PT}}$-symmetric quantum theory, it turned out that ${\cal {PT}}$-symmetric quantum theory
admitted 
states which have negative norms. This had no clear interpretation. 
This was cured by introducing another operator ${\cal C}$ called as the conjugation 
operator \cite{bend1,bend2}. This operator commutes with the Hamiltonian 
and the operator ${\cal {PT}}$. Also, note that ${\cal C}^2=I $, which implies that it has 
eigenvalues $\pm 1$.

Bender {\it et al} \cite{bend1,bend2} have shown that non-Hermitian 
Hamiltonians can have real eigenvalues if it possess 
${\cal {PT}}$-symmetry, i.e., $[H, {\cal PT}]= 0$  and the symmetry is unbroken 
(if all of the eigenfunctions of $H$  are simultaneous eigenfunction of 
the operator ${\cal {PT}}$). Hamiltonians having unbroken ${\cal {PT}}$ symmetry can 
define a unitary quantum theory.
Unitarity can be shown by the fact that such Hamiltonians possess a new 
symmetry called conjugation ${\cal C}$ with $[{\cal C}, H]=0$ and $[{\cal C}, {\cal PT}]=0$.

Quantum theory that deals with non-Hermitian Hamiltonians and 
respects ${\cal {CPT}}$ symmetry may be called non-Hermitian 
quantum theory. One can formalize the framework by stating the following postulates:
%\noindent
(i) A quantum system is a three-tuple $({\cal H}, H,  
\langle . | . \rangle_{{\cal CPT}} )$, where ${\cal H}$ is a physical Hilbert 
space with the ${\cal {CPT}}$ inner product $\langle . | . \rangle_{{\cal CPT}}$ having a 
positive norm, and $H$ is the non-Hermitian Hamiltonian,
%\noindent
 (ii) The state of a system is a vector $|\psi \rangle$ in ${\cal H}$.
 For any two vectors the ${\cal {CPT}}$ inner product is defined as
 $\langle \psi | \phi \rangle_{{\cal CPT}} = 
\int~ dx [ {\cal {CPT}} \psi(x)] \phi(x)$, 
%\noindent
(iii) The time evolution of state vector is unitary with respect to ${\cal {CPT}}$
inner product, 
%\noindent
(iv) An observable can be a linear operator $O$, provided it is Hermitian 
with respect to the ${\cal {CPT}}$ inner product, i.e., 
$\la .|O~. \ra_{\cal {CPT}} = \la O~ .| .\ra_{\cal {CPT}} $, 
%\noindent
(v) If we measure an observable $O$, then the eigenvalues are the possible 
outcomes, 
%\noindent
(vi) If measurement gives an eigenvalue $O_n$, the states makes a transition 
to the eigenstate $|\psi_n\ra$ and the probability of obtaining the 
eigenvalues $O_n$ (say) in a state 
$|\psi\rangle$ is given by 
\begin{eqnarray}
p_n =\frac{ |\langle \psi | \psi_n \rangle_{\cal {CPT}} |^2 }
{||\psi||_{{\cal {CPT}}} ||\psi_n||_{{\cal {CPT}} }},
\end{eqnarray}
where $||\psi ||_{{\cal {CPT}}} = \sqrt{\langle \psi | 
\psi \rangle_{{\cal  {CPT}}} }$,
and 
%\noindent
(vii) If we have two quantum systems $({\cal H}_1, H_1,  
\langle . | . \rangle_{\cal {CPT}} )$ and 
$({\cal H}_2, H_2,  
\langle . | . \rangle_{\cal {CPT}} )$, then the state of the combined system 
lives in a tensor product Hilbert space ${\cal H}_1 \otimes {\cal H}_2$.\\

{\em Entanglement for ${\cal {PT}}$ symmetric qubits.--}
In ${\cal {PT}}$-symmetric quantum mechanics if
we store information in any two distinct orthogonal states of non-Hermitian Hamiltonian, 
then we call it as a 
${\cal {PT}}$-symmetric quantum bit or in short a {\em PTqubit}.
In general a ${\cal {PT}}$qubit is different from a qubit. In the limit of 
vanishing non-Hermiticity parameter, a ${\cal {PT}}$qubit becomes a 
standard qubit.

In non-Hermitian quantum theory a general two-state system 
will be described by a $2\times 2$ Hamiltonian which respects ${\cal {CPT}}$ symmetry.
Following Bender {\it et al} \cite{bend1}, this Hamiltonian is given by

\begin{eqnarray}
%\begin{displaymath}
H = \left( \begin{array}{rr} r~e^{i\theta} & s \\  t~~ &   r~e^{-i\theta} 
\end{array} \right),
%\end{displaymath}
\end{eqnarray}
with $r, s, t,$ and $\theta$ all real numbers. It has eigenvalues 
$E_{\pm} = r \cos \theta \pm \sqrt{st -r^2 \sin^2 \theta}$.
This Hamiltonian is non-Hermitian yet it has real eigenvalues whenever we have $st > r^2 \sin^2 
\theta$. Also, $H$ is invariant under ${\cal {CPT}}$. Two distinct 
eigenstates of this Hamiltonian are given by 
\begin{eqnarray}
%\begin{displaymath}
|\psi_+\ra = \frac{1}{\sqrt{ 2 \cos \alpha}} 
\left( \begin{array}{r} e^{i\alpha/2} \\  e^{-i\alpha/2} \end{array} \right)~~ 
|\psi_-\ra = \frac{1}{\sqrt{ 2 \cos \alpha}} 
\left( \begin{array}{r} e^{-i\alpha/2} \\  -e^{i\alpha/2}  
\end{array} \right), \nonumber\\
%\end{displaymath}
\end{eqnarray}
where $\alpha$ is defined through $\sin \alpha = \frac{r}{\sqrt{st}} 
\sin \theta$. With respect to the ${\cal {CPT}}$ inner product (which gives a positive 
definite inner product) we have 
$\la \psi_{\pm} | \psi_{\pm} \ra_{\cal CPT} = 1$ and 
$\la \psi_{\pm} | \psi_{\mp} \ra_{\cal CPT} = 0$. The ${\cal {CPT}}$ inner product for 
any two states of ${\cal {PT}}$Qubit is given by
\begin{eqnarray}
\la \psi|\phi \ra = [({\cal CPT}) |\psi \ra]. \phi, 
\end{eqnarray}
where $\la \psi|$ is the ${\cal {CPT}}$ conjugate of $|\psi\ra$.
In the $2$-dimensional Hilbert space, the operator $C$ is given by 
\begin{eqnarray}
C = \frac{1}{\cos \alpha} \left 
( \begin{array}{rr} i\sin \alpha & 1 \\  1  &  -i\sin \alpha  
\end{array} \right).
\end{eqnarray}
The operator ${\cal P}$ is unitary and is given by
%\begin{eqnarray}
${\cal P} =  \left ( \begin{array}{rr} 0 & 1 \\  1  &  0 \end{array} \right)$.
%\end{eqnarray}
The operator ${\cal T}$ is anti-unitary and its effect is to transform 
$x \rightarrow x, p \rightarrow -p$ and
$i \rightarrow -i$.  

Since the eigenstates $|\psi_{\pm}\rangle$ of the non-Hermitian Hamiltonian 
$H$ span the 
two-dimensional Hilbert space, one can encode
one bit of information in these orthogonal states. 
An arbitrary state can be 
represented as superposition of these orthogonal states 
\begin{eqnarray}
|\Psi \ra = \alpha |\psi_+ \ra + \beta |\psi_-\ra = 
\alpha |0_{\cal {CPT}} \ra + \beta |1_{\cal {CPT}}\ra.
\end{eqnarray}
Thus, any arbitrary 
superposition of two orthogonal states of ${\cal {PT}}$ invariant Hamiltonian will be
called ${\cal {PT}}$-quantum bit or {\it PTqubit}. In fact, any linear superposition 
of two orthogonal states of an observable $O$ in ${\cal {PT}}$-symmetric quantum 
theory can represent a {\it PTqubit}.

In ${\cal {PT}}$-symmetric quantum theory quantum entanglement can arise if we  
have more than one ${\cal {PT}}$qubit. Now, 
suppose we have two quantum systems with non-Hermitian 
Hamiltonians $H_1$ and $H_2$, where
\begin{eqnarray}
H_1 = \left( \begin{array}{rr} r e^{i\theta} & s \\  s~~ &   r e^{-i\theta} 
\end{array} \right)~ 
H_2 = \left( \begin{array}{rr} r'e^{i\theta'} & s' \\  s'~~ &   
r'e^{-i\theta'} \end{array} \right).
\end{eqnarray}
Let  $\{ |\psi_{\pm}\rangle \} \in {\cal H}_1$ and 
$\{ |\psi'_{\pm}\rangle \} \in {\cal H}_2$ are the eigenfunctions of the 
Hamiltonians $H_1$ and $H_2$, respectively.
The state of the combined system will live in ${\cal H}_1 \otimes 
{\cal H}_2$ which is spanned by $\{ |\psi_+\rangle \otimes 
|\psi_+'\rangle, |\psi_+\rangle \otimes |\psi_-'\rangle, |\psi_- \rangle 
\otimes |\psi_+'\rangle, |\psi_-\rangle \otimes |\psi_-'\rangle \}$. 
%If the combined state cannot be written as 
%$|\Psi\rangle = |\psi\rangle \otimes |\phi\rangle = 
%|\psi\rangle |\phi\rangle $, then it is entangled.
A general state of two ${\cal {PT}}$qubits can be expanded using the joint basis 
in ${\cal H}_1 \otimes {\cal H}_2$ and it will be an entangled state.
%\begin{eqnarray}
%|\Psi\rangle = a |\psi_+\rangle \otimes |\psi_+'\rangle 
% + b |\psi_+\rangle \otimes |\psi_-'\rangle + c |\psi_-\rangle 
%\otimes |\psi_+'\rangle
%+ d |\psi_-\rangle \otimes |\psi_-'\rangle.
%\end{eqnarray}

The ${\cal {CPT}}$ inner products on the Hilbert spaces ${\cal H}_1$ and 
${\cal H}_2$ induce the inner product on 
${\cal H}_1 \otimes {\cal H}_2$.
For any two arbitrary vectors $|\Psi\rangle, |\Phi \rangle 
\in {\cal H}_1 \otimes {\cal H}_2$, we define the inner product between them 
as   

\begin{eqnarray}
\langle \Psi | \Phi \rangle_{{\cal CPT}} = 
[({\cal {CPT}}) \otimes ({\cal {CPT}} ) |\Psi\rangle]. |\Phi \rangle.
\end{eqnarray}
Using this inner product we can calculate relevant physical quantities for 
the composite system under consideration.

Now, we come to the central question: is it possible to describe a composite system where one
part is described by a local ${\cal {PT}}$  symmetric Hamiltonian and the other part is by conventional 
quantum theory. We will show that this will lead to contradiction with a basic law of 
quantum entanglement.

{\em Violation of Entanglement Invariance Under Local Unitary.--}
For any pure bipartite state $|\Psi\ra_{AB}$ the entropy of any one of the reduced density matrix 
is a measure of entanglement \cite{pr}.
It is given by

\begin{eqnarray}
E(\Psi ) = 
- \mbox{tr}_A (\rho_A \log \rho_A) =  - \mbox{tr}_B(\rho_B \log \rho_B),
%= - \sum_i p_i \log p_i.
\end{eqnarray}
where $\rho_A = \mbox{tr}_B (|\Psi\ra_{AB} \la \Psi |) $ and $\rho_B = \mbox{tr}_A (|\Psi\ra_{AB} \la \Psi |)$.
This measure of entanglement satisfies the following properties \cite{plenio}:
(i) $E(\Psi) = 0$ iff $|\Psi \ra$ is separable, 
(ii) $E(\Psi)$ is invariant under local unitary transformations, i.e., 
$E(\Psi) = E(U_1 \otimes V_2 \Psi)$, 
(iii) $E(\Psi)$ cannot increase under local operation and classical 
communications (LOCC)
and (iv) the entanglement content of $n$ copies of $|\Psi\ra$ is additive, i.e., 
$E(\Psi^{\otimes n} ) =  n E(\Psi)$.

Consider a situation where Alice and Bob share maximally entangled state as described by
conventional quantum theory, i.e.,  a state of the form 

\begin{eqnarray}
|\Phi \rangle = \frac{1}{\sqrt 2}(|0  \rangle
| 0 \rangle + |1 \rangle | 1 \rangle,
\end{eqnarray}
where $|0\ra, |1\ra$ are the eigenstates of $\sigma_z$.
Suppose Alice has a locally ${\cal {PT}}$ symmetric quantum system with a Hamiltonian $H$ and 
applies a local unitary $U(t)= \exp(-itH)$ ($\hbar =1$) to her subsystem.
Then the composite system will evolve as
\begin{eqnarray}
|\Phi(t) \rangle = 
%= e^{-iHt} \otimes I)|\Psi \rangle = 
\frac{1}{\sqrt 2}(e^{-iHt}|0  \rangle | 0 \rangle + e^{-iHt}|1 \rangle | 1 \rangle.
\end{eqnarray}
Using the resolution of identity $\sum_{n=\pm} |\psi_n \ra \la \psi_n| = I$ for the eigenstates of 
the ${\cal {PT}}$ symmetric 
Hamiltonian, we can write the above time-evolved state as 
\begin{eqnarray}
|\Phi(t) \rangle = \frac{1}{\sqrt 2}(e^{-iE_+t}|\psi_+  \rangle
| \phi_+ \rangle + e^{-iE_-t}|\psi_- \rangle | \phi_- \rangle.
\end{eqnarray}
where $| \phi_+ \rangle = c_+^{(0)} | 0 \rangle + c_+^{(1)}| 1 \rangle$,
 $| \phi_- \rangle = c_-^{(0)} | 0 \rangle + c_-^{(1)}| 1 \rangle$, 
$c_+^{(0)} =\la \psi_+|0\ra, c_+^{(1)} =\la \psi_+|1\ra,  
c_-^{(0)} =\la \psi_-|0\ra$, and $ c_-^{(1)} = \la \psi_-|1\ra$. Note that
here $\la \psi_{\pm}|$ are ${\cal {CPT}}$ conjugates of $|\psi_{\pm} \ra$.
One can simplify the above by noting that $c_+^{(0)} = c, c_+^{(1)}= c^*, c_-^{(0)}=c^*$,
and $c_-^{(1)} = -c$, where $c= \frac{e^{i\alpha/2}}{\sqrt{2 \cos \alpha}}$.

Since Bob is in the conventional quantum world, he will apply the rules of standard quantum theory.
After the action of ${\cal {PT}}$ symmetric local unitary on Alice's subsystem, the reduced state
of Bob is given by  (with normalization)  
%${\tilde \rho}_1 = \rho_1/\mbox{Tr } \rho_1 $, so that 
\begin{eqnarray}
\label{den}
\rho_B &=& \frac{1}{N} 
( 2 + \cos (Et-2\alpha) -\cos Et)|0\ra \la 0|  \nonumber\\
&+& 2i \sin \alpha (1-\cos Et) |0\ra \la1|   
-  2i \sin \alpha (1-\cos Et) |1\ra \la 0| \nonumber\\
&+&  ( 2 + \cos (Et + 2\alpha) -\cos Et)|1\ra \la 1| .
\end{eqnarray}
where $E= (E_+ -E_-)$ and $N = 4(1- \cos Et \sin^2\alpha)$. The eigenvalues of the reduced 
density operator are simply $\lambda_{\pm}(t) = \frac{1}{2} \pm K(t)$, where $K(t)$ is given by
\begin{eqnarray}
K(t) = \frac{(7 -8 \cos Et + \cos 2Et + 2 \cos 2\alpha \sin^2 Et)^{\frac{1}{2}} \sin \alpha}{4(1- \cos Et \sin^2\alpha)}. \nonumber\\
\end{eqnarray}
Therefore, the entanglement of the state $|\Phi(t)\ra$ is given by $E(\Phi(t)) = -\sum_{i=\pm} \lambda_i(t) 
\log \lambda_{i}(t)$, which is not unity.
Thus, by a local ${\cal {PT}}$ symmetric unitary transformation, the maximally entangled state
has changed to a non-maximally entangled state, there by showing that entanglement 
is not preserved under such local unitary. For simplicity, if we take $Et=\pi/2$, then we can see
that the reduced density matrix is given by

\begin{eqnarray}
{ \rho}_B = \frac{1}{2} 
\left 
( \begin{array}{rr} 1+ \sin \alpha \cos \alpha ~~~  & i \sin \alpha \\  
-i \sin \alpha~~~~~~~~  &   1- \sin \alpha \cos \alpha
\end{array} \right)
\end{eqnarray}
which depends only on the non-Hermiticity parameter $\alpha$.
The eigenvalues of the density matrix ${ \rho}_B$ are given by
$\lambda_{\pm} = \frac{1}{2}(1 \pm  \sqrt{(1- \cos^4 \alpha)}$. The violation of conservation of entanglement
under local ${\cal {PT}}$ symmetric unitary transformation holds for all values of
the non-Hermitian parameter $\alpha$. When $\alpha=0$, we see that entanglement is preserved
under local unitary.
This shows that under a ${\cal {PT}}$ symmetric local unitary transformation on one part, 
the entanglement as seen by an observer in the conventional quantum world is not preserved. 
A local ${\cal {PT}}$ symmetric unitary transformation
acts like a global operation on the two entangled particles. Since this violates a 
basic conservation law of entanglement, most likely option may be that a local 
${\cal {PT}}$ symmetric unitary transformation should not coexist with the conventional
quantum theory.

One can see that the recent result that local ${\cal {PT}}$ symmetry violates the no-signaling condition \cite{lee} 
also follows from our result.
Suppose that Alice want to send one classical bit using the shared maximally entangled state. If She wishes to send $`0'$, 
she decides to do nothing and if she wants to send $`1'$ she applies a local ${\cal {PT}}$ symmetric 
unitary. Accordingly, the state of Bob will be different: in the first case it is a random mixture $\frac{I}{2}$ and in the 
second case it is given by (\ref{den}). 
Thus, by a local ${\cal {PT}}$ symmetric unitary transformation, the state of Bob has changed
from a random mixture to a non-degenerate density operator $\rho_B$.
Since the local action by Alice changes the reduced density matrix for 
Bob who may be space-like separated, this can lead to signaling.

{\em Entanglement mismatch and signaling.---}
Here, we will illustrate the effect of non-Hermiticity on the entanglement, which we call `entanglement mismatch'
in switching from usual quantum theory to ${\cal {PT}}$ symmetric quantum theory or vice versa. 
We will argue that this effect is already a signature of the violation of the no-signaling condition.
First, note that the partial trace operation is indeed a quantum operation. Let us define the Kraus 
elements for the partial trace over the subsystem $A$ as $E_i = \la \psi_i | \otimes I$, so that we have $E_i^{\dagger} = 
|\psi_i \ra \otimes I$ for some orthonormal basis $|\psi_i\ra \in {\cal H}_A$. This satisfies $\sum_iE_i^{\dagger}  E_i =I$.
If we have a composite state $\rho_{AB}$, then under this quantum operation we have $\rho_{AB} 
\rightarrow \sum_i  E_i \rho_{AB} E_i^{\dagger} =  \mbox{tr}_A (\rho_{AB}) = \rho_B$. 
Moreover a local operation on the subsystem $A$ cannot change 
the reduced state of the subsystem $B$. However, we will show that such a local operation in ${\cal {PT}}$
symmetric world can change the reduced state of the other subsystem which may be space-like separated.

Suppose that Alice and Bob share an entangled state
%\begin{eqnarray}
$|\Psi \rangle_{AB} = \sum_{i} \sqrt{\lambda_i} 
|\psi_i \rangle_A \otimes |\phi_i \rangle_B$,
%\end{eqnarray}
where $\lambda_i$'s are the Schmidt coefficients, and $|\psi_i \rangle$, $|\phi_i \rangle$ are the Schmidt basis.
If Alice and Bob both describe their particles using conventional quantum theory, then the state of 
Bob's particle is $\rho_B = \sum_{i} \lambda_i |\phi_i \rangle \langle \phi_i |$. The entanglement 
entropy is given by $E(\Psi) = -\sum_{i} \lambda_i \log \lambda_i$. Now, 
we will show that if Alice is in a ${\cal {PT}}$ symmetric quantum world, then 
the reduced states of the particle $B$ will be 
different. 
%If we calculate the partial traces in usual quantum theory and  in non-Hermitian quantum theory.
Because the inner products in ordinary and ${\cal {PT}}$-symmetric quantum theory 
are different, the partial traces will give different results. For example, if Alice is in a 
${\cal {PT}}$ symmetric quantum world, 
the reduced density matrix for the particle $B$ will be 
\begin{eqnarray}
\rho_B
&=& \sum_{ij} \sqrt{\lambda_i \lambda_j } 
[({\cal {CPT}}) |\psi_j \rangle] . |\psi_i \rangle |\phi_i \rangle \langle \phi_j |   \nonumber\\
&=& \sum_{ij} \sqrt{\lambda_i \lambda_j }  \la \psi_j |\psi_i \rangle
|\phi_i \rangle \langle \phi_j |.
\end{eqnarray}
The above density matrix is not in the diagonal form because 
$\la \psi_j |\psi_i \ra \not= \delta_{ij}$ in the usual sense. 
%Similarly, one can check that  the reduced density matrix of the particle $B$ will be different in two theories. 
As a consequence, the entanglement content of a bipartite state depends on the 
observer's world.
This phenomenon we call as the `entanglement mismatch' in switching from 
conventional quantum world to ${\cal {PT}}$ symmetric quantum world.

To see this clearly, let us consider the situation where Alice and Bob share an entangled state of spin-singlet in 
ordinary quantum theory. If both observers are in the conventional quantum world, then they will agree that they share 
a maximally entangled state. However, if Alice is in a ${\cal {PT}}$ symmetric quantum world, the reduced 
density matrix for particle $B$ is given by 
%\begin{eqnarray}
$\rho_B = \mbox{tr}_A(|\Psi^- \ra \la \Psi^- |) = \frac{1}{2} 
[|0\ra \la 0| \la 1|1\ra_{\cal {CPT}}  
- |0\ra \la 1| \la 0|1\ra_{\cal {CPT}} 
- |1 \ra \la 0 | \la 1|0\ra_{\cal {CPT}}  + |1 \ra \la 1| \la 0|0\ra_{\cal {CPT}} ]$, 
%\nonumber \\
%\end{eqnarray}
where the ${\cal {CPT}}$ inner products are given by 
$\langle 0 |0 \rangle_{\cal {CPT}} = \langle 1|1\rangle_{\cal {CPT}}  
= \frac{1}{\cos \alpha}, 
\langle 0 |1 \rangle_{\cal {CPT}} = i \tan \alpha$ and  
$\langle 1 |0 \rangle_{\cal {CPT}} 
= -i \tan \alpha$. Using these, the reduced density matrix $\rho_B$ 
is given by (after renormalization)
%\begin{eqnarray}
%\rho_1 = \mbox{tr}_2(|\Psi^- \rangle \langle \Psi^- |) = 
%\frac{1}{2 \cos^2 \alpha} 
%\left 
%( \begin{array}{rr} 1+\sin^2\alpha~  & -2i \sin \alpha \\  
%2i \sin \alpha~  &  1 +\sin^2\alpha
%\end{array} \right).
%\end{eqnarray}
%Note that $\rho_1$ is not normalized. We can define a normalized density 
%matrix ${\tilde \rho}_1 = \rho_1/\mbox{Tr } \rho_1 $, so that 
\begin{eqnarray}
{ \rho}_B = \frac{1}{2} 
\left 
( \begin{array}{rr} 1~~~  & -i \sin \alpha \\  
i \sin \alpha  &  1 
\end{array} \right).
\end{eqnarray}
The eigenvalues of the density matrix $\rho_B$ are now given by
$\lambda_{\pm} = \frac{1}{2}(1 \pm  \sin \alpha)$
%~~~ \lambda_2 =  \frac{1}{2}(1 - 2 \sin \alpha)$.
Therefore, the entanglement entropy 
is given by 
%\begin{eqnarray}
$E(\Psi^-) 
%&=& -\lambda_1 \log \lambda_1 - \lambda_2 \log \lambda_2 
=  - \frac{1}{2}(1 + \sin \alpha) \log \frac{1}{2}(1 +  \sin \alpha) 
%\nonumber \\
- \frac{1}{2}(1 -  \sin \alpha) \log \frac{1}{2}(1 -  \sin \alpha)$.
%\end{eqnarray}
In the Hermitian limit ($\alpha =0$), Alice and Bob will share an entangled state with $E(\Psi^-) = 1$.
This shows that if Alice is located in conventional quantum world then two distant parties will share one 
unit of entanglement, however if she is having access to ${\cal {PT}}$ symmetric quantum world, the shared state 
will have less than one unit of entanglement. 
This is the ``entanglement mismatch'' effect which arise dues to non-Hermiticity. 
%It appears that if we remain in the same world we cannot see the effect 
%of non-Hermiticity.

%Similarly, if we define a maximally entangled state 
%$|\Psi_{\cal CPT} \ra = \frac{1}{2}(|\psi_+ \ra |\psi_-\ra - |\psi_- \ra 
%|\psi_+\ra) $ in ${\cal {PT}}$-symmetric quantum theory, then we will have less than one unit of 
%entanglement in ordinary quantum theory. 

 To see that the ``entanglement mismatch'' indeed leads to signaling, note that the state of Bob's particle
 changes depending on the situation whether Alice is in conventional quantum world or in ${\cal {PT}}$ symmetric quantum world. 
 Since a local operation should not change the reduced state of a remote particle, this violates the 
 no-signaling condition.  One can also understand the signaling by saying that Alice carries out measurements in conventional 
 computational basis $\{|0\ra, |1\ra\}$ or in ${\cal {PT}}$ symmetric basis $\{|\psi_+\ra, |\psi_-\ra\}$. Depending on this
 choice, the reduced state of Bob will be different which can in principle be distinguishable. 

%{\em Signaling with local ${\cal {PT}}$ symmetric operation.--}

{\em Outlooks.--}
Quantum theory has phenomenal predictive power and still it continues to predict new effects. Can we say similar 
things for the complex extension of quantum theory governed by ${\cal {PT}}$ symmetric Hamiltonians? To answer such 
questions, one must go beyond the single particle description of non-Hermitian Hamiltonian systems to composite 
systems. Indeed, we have shown that the notion of ${\cal {PT}}$qubit and entanglement for composite quantum systems described by 
${\cal {PT}}$ symmetric Hamiltonians can be introduced in a consistent manner.
%However, a qubit which is in orthogonal state in ${\cal {PT}}$-symmetric quantum theory  
%becomes non-orthogonal in ordinary quantum theory and vice verse.  This has several consequences in quantum information. 
However, a pressing question is whether it is possible to describe a composite system where one
part is described by a local ${\cal {PT}}$  symmetric Hamiltonian and the other part is by conventional 
quantum theory, and whether it is possible to switch between two quantum worlds.  
We have shown that this, in fact, leads to a contradiction with a basic law of conservation of 
quantum entanglement.  Specifically, we have shown that an observer having access to ${\cal {PT}}$ symmetric local unitary,
then that 
can change the entanglement content of the shared resource. Thus, one of the fundamental 
law of quantum entanglement is violated by local ${\cal {PT}}$ symmetric unitary. 
We have shown how the entanglement property of quantum states also
change if we switch from usual quantum world to non-Hermitian world. If one of the particle is in a 
${\cal {PT}}$ symmetric quantum world, then a maximally entangled state in the sense of 
ordinary quantum theory will appear as  non-maximally entangled state for another observer. 
This shows that it is not enough to share a maximally entangled state between two distant parties, 
but they should know in which world 
%the maximally entangled state has been prepared and in which world 
they are located. This effect we call as the ``entanglement mismatch'' and this can be regarded as a signature of 
the violation of the no-signaling condition. 

Therefore, one may accept that the complex extension of quantum theory with ${\cal {PT}}$ symmetric
Hamiltonian may be a genuine extension of quantum theory, as this can have inequivalent predictions 
for spatially separated systems. If this is true, then ${\cal {PT}}$ symmetric Hamiltonians may be 
used as a powerful resource for quantum computation, quantum information and quantum communications. 
Other option is that local ${\cal {PT}}$ symmetry may not coexist with 
conventional quantum theory and it may not be 
implementable in physics. 
%Only future experiment can decide which way we should go.
%as this leads to the violation of the conservation of entanglement.

\vskip .8cm

\noindent
{\it Acknowledgement:} I thank Uttam Singh for useful discussions.

\renewcommand{\baselinestretch}{1}

\end{document}